\documentstyle[titlepage,12pt]{article}
\def\thefootnote{\fnsymbol{footnote}}
\def\@makefnmark{\mbox{$^\@thefnmark$}}
\def\thefootnote{\mbox{\noindent$\fnsymbol{footnote}$}}
\long\def\@makefntext#1{\noindent$^{\@thefnmark}$#1}

\def\appendix{%\vskip 1cm\par\setcounter{equation}{0}
\def\theequation{A1.\arabic{equation}}}

\def\openone{\leavevmode\hbox{\small1\kern-3.3pt\normalsize1}}

\newcommand{\preprint}[2]{\large
\begin{center}
\hspace*{8cm} Preprint  \, EFUAZ\\
\hspace*{8cm} FT-94-{#1}\\
\hspace*{8cm} hep-th/9410174\\
\hspace*{8cm}  {#2} 1994
\end{center}}

\begin{document}

\title{\vspace*{-35mm} \preprint{09-REV}{October} \vspace*{2cm}
{\Large {\bf Can the $2(2S+1)$ Component Weinberg-Tucker-Hammer
Equations Describe  the~Electromagnetic
Field?}}\thanks{Submitted to ``Phys. Rev. Lett."}}

\author{{\bf Valeri V. Dvoeglazov}\thanks{On leave of absence from
{\it Dept. Theor. \& Nucl. Phys., Saratov State University,
Astrakhanskaya str., 83, Saratov\, RUSSIA}}\,
\thanks{Email: valeri@bufa.reduaz.mx,
dvoeglazov@main1.jinr.dubna.su}\\
%{\em
Escuela de F\'{\i}sica, Universidad Aut\'onoma de Zacatecas \\
Antonio Doval\'{\i} Jaime\, s/n, Zacatecas 98000, ZAC., M\'exico}
%}

\begin{abstract}

It is shown that the massless $j=1$ Weinberg-Tucker-Hammer
equations reduce to the Maxwell's equations for
electromagnetic field under the definite choice of
field functions and initial and boundary conditions.
Thus, the former appear to be of use in
a description of some physical processes for
which that could be necessitated or be convenient.

The possible consequences are discussed.
\end{abstract}

\date{\empty}

\maketitle

\renewcommand{\thefootnote}{\arabic{footnote}}

\newpage

\normalsize{
The attractive Weinberg's $2(2j+1)$ component formalism for a description
of higher spin particles~\cite{Weinberg} recently got developed
considerably in connection with the recent works of
Dr.~D.V. Ahluwalia {\it et al.}, ref.~\cite{DVA1}-\cite{DVA4}.
Some attempts have also been done in attaching the interpretation of their
ideas in my papers~\cite{DVO1}-\cite{DVO4}.

The main equation in this formalism,
which had been proposed in ref.~\cite{Weinberg},
is\footnote{For discussions on the needed
modifications of the theory see refs.~\cite{DVA3}-\cite{DVO3}
and what follows.}
\begin{equation}\label{eq0}
\left [\gamma_{\mu_1 \mu_2 \ldots \mu_{2j}} p_{\mu_1} p_{\mu_2} \ldots
p_{\mu_{2j}}+ m^{2j}\right ] \psi = 0,
\end{equation}
One can see that it is of the ``$2j$" order in the momentum,
$p_{\mu_i}=-i\partial/\partial x_{\mu_i}$, $m$ is a particle mass. The
analogues of the Dirac $\gamma$- matrices are $2(2j+1)\otimes 2(2j+1)$
matrices which have also ``$2j$" vectorial indices, ref.~\cite{Barut}.

For the moment I take a liberty to repeat the previous results.
The equations (4.19,4.20), or equivalent to them Eqs. (4.21,4.22),
presented in ref.~[1b,p.B888] and in many other publications:
\begin{eqnarray}
{\bf \nabla}\times \left [ {\bf E} -i {\bf B}\right ]
+i (\partial/\partial t) \left [{\bf E} -i{\bf B} \right ] &=&0, \qquad
(4.21) \nonumber\\
{\bf \nabla}\times \left [ {\bf E} +i {\bf B}\right ]
-i (\partial/\partial t) \left [{\bf E} +i{\bf B} \right ] &=&0, \qquad
(4.22) \nonumber
\end{eqnarray}
are found in ref.~\cite{DVA1} to have acausal
solutions.  Apart from the correct dispersion relations $E=\pm p$ one has
a wrong dispersion relation $E=0$. The origin of this fact seems to
be the same to
the problem of the ``relativistic cockroach nest" of
Moshinsky and Del Sol, ref.~\cite{Moshinsky}.
On the other hand, {\it "the $m\rightarrow 0$
limit of Joos-Weinberg finite-mass wave equations, Eq. (\ref{eq0}),
satisfied by $(j,0)\oplus (0,j)$ covariant spinors, ref.~\cite{DVA2},
are free from all kinematic acausality."}

The same authors (D. V. Ahluwalia and his
collaborators) proposed
the Foldy-Nigam-Bargmann-Wightman-Wigner-type (FNBWW) quantum field
theory, {\it ``in which bosons and antibosons  have
opposite relative intrinsic parities"}, ref.~\cite{DVA3}.
This {\it Dirac}-like modification of the Weinberg theory
is an excellent example of combining the Lorentz and the
dual transformations. Its recent development, ref.~\cite{DVA4},
could be relevant for a description of neutrino oscillations
and in realizing the role of space-time symmetries for all types
of interactions.

In ref.~\cite{DVO1} I concern with a connection between
antisymmetric tensor fields~\cite{AVD1,AVD2} and
the equations considered by Weinberg (and by Hammer and Tucker~\cite{Hammer}
in a slightly different form). In the case of the choice\footnote{My
earlier attempts to give a potential interpretation of the $\psi$ were
unsuccessful in a certain manner, ref.~\cite{OLD}.}
\begin{equation}\label{psi}
\psi=\pmatrix{{\bf E}+i{\bf B} \cr {\bf E}-i{\bf B}\cr}
\end{equation}
the equivalence of the Weinberg-Tucker-Hammer approach and the Proca
approach\footnote{I mean the equations for $F_{\mu\nu}$,
the antisymmetric field tensor, e.~g., Eq. (3) in~\cite{DVO1}
or Eq. (A4) in~[16b].} has been found.
The possibility of a consideration of another
equation (the Weinberg ``double")\footnote{It is useful to compare
the method applied in the papers~\cite{DVA1,DVO1}
with the Dirac's way of deriving the famous equation for $j=1/2$
particles, ref.~\cite{Dirac}.
Namely, his aim was to obtain the linear differential
equation; the coefficients at derivatives and a mass term
were not known {\it ab initio} and they turned out
to be matrices. The second requirement which he imposed is:
the equation should be compatible with the Klein-Gordon
equation, i.~e. with relativistic dispersion relations.}
was point out. In fact, it is the equation for the
antisymmetric tensor dual to $F_{\mu\nu}$,
which had also been considered earlier,
e.~g.~ref.~\cite{Takahashi}.  In the paper~\cite{DVO2} the Weinberg fields
were shown to describe the particle with transversal components (i.~e.,
spin~$j=1$) as opposed to the conclusions of refs.~\cite{AVD1,AVD2} and of
the previous ones~\cite{Hayashi}.  The origins of contradictions with the
Weinberg
theorem ($B-A=\lambda$), which have been met in the old works (of both
mine and others), have been partly clarified\footnote{Answering to the referee
of one of my previous paper {\it I can ascertain} my statements. The Weinberg
theorem is a consequence of the {\it general} kinematical structure of the
theory
based on the definite {\it representation of the Lorentz group}. If  one
of {\it the various ``gauge" constraints  one  places on the dynamics} leads
to the results which contradict with the underlying kinematical structures
this can signify the only thing: the dynamical constraint is wrong! {\it This
is
a known result and reproducing  it  once again}  is caused by the importance
of {\it this particular example.} Namely, the Weinberg theorem permits
two values of the helicity $\lambda=\pm 1$ for a massless $j=1$
Weinberg-Tucker-Hammer
field. Setting the generalized Lorentz condition (see for a discussion
the footnote \# 10 in ref.~\cite{DVO2}) yields the {\it physical
excitation} of the very strange {\it nature}, $\lambda =0$, which also
contradicts
with a classical limit, see ref.~\cite{DVO1}.
Therefore,  imposing the generalized Lorentz condition, formulas (18)
of ref.~\cite{Hayashi}, on the {\it quantal} physical states is impossible
in the case of a quantum field consideration and it is incompatible with
{\it the specific properties of an antisymmetric tensor field.} }
The causal propagator of the Weinberg theory has been proposed in
ref.~\cite{DVO3}. Its remarkable feature is a presence of
four terms. This fact is explained in my forthcoming paper~\cite{DVO4}.

The aim of the present Letter is to  consider the question,
under what conditions the Weinberg-Tucker-Hammer
$j=1$ equations can be transformed to Eqs. (4.21) and (4.22)
of ref.~[1b]~? By using the
interpretation of $\psi$ in the chiral representation,
Eq. (\ref{psi}), and the explicit form of
the Barut-Muzinich-Williams matrices, ref.~\cite{Barut},
I am able to recast the $j=1$ Tucker-Hammer equation
\begin{equation}
\left [\gamma_{\mu\nu} p_\mu p_\nu +p_\mu p_\mu +2m^2 \right ] \psi = 0,
\end{equation}
which is free of tachyonic solutions,
or the Proca equation, Eq. (3) in ref.~\cite{DVO1}, to the
form\footnote{I restored  $c$, the light velocity, at the
terms.}$^,$\footnote{The reader can reveal dual equations
from  Eqs. (10) or (12) and parity-conjugated equations
from Eqs. (18,19)  of ref.~\cite{DVO1} without any problems.}
\begin{eqnarray}\label{ME0}
m^2 E_i &=& -{1\over c^2} {\partial^2 E_i \over \partial t^2}
+\epsilon_{ijk} {\partial \over \partial x_j}
{\partial B_k \over \partial t} +
{\partial \over \partial x_i} {\partial E_j \over \partial x_j},\\
\label{ME1}
m^2 B_i &=& {1\over c^2} \epsilon_{ijk} {\partial \over
\partial x_j} {\partial E_k \over \partial t} +  {\partial^2 B_i \over
\partial x_i^2} -{\partial \over \partial x_i} {\partial B_j \over
\partial x_j}.
\end{eqnarray}
The D'Alembert equation (the Klein-Gordon equation in
the momentum representation indeed)
\begin{equation}\label{DA}
\left ({1\over c^2} \frac{\partial^2}{\partial t^2} -
\frac{\partial^2}{\partial x_i^2}\right ) F_{\mu\nu} = - m^2 F_{\mu\nu}
\end{equation}
is implied.

Restricting ourselves by the consideration of
the $j=1$ massless case one can
re-write them to the following form:
\begin{eqnarray}\label{MY1}
{\partial \over \partial t}\, curl\, {\bf B} + grad\, div\, {\bf E}
-{1\over c^2} {\partial^2 {\bf E} \over \partial t^2} &=& 0,\\
\label{MY2}
{\bf \nabla}^2 {\bf B} -grad\, div\, {\bf B} +{1\over c^2} {\partial \over
\partial t}\, curl\, {\bf E} &=& 0.
\end{eqnarray}

Let consider the first equation (\ref{MY1}). We can satisfy it
provided that
\begin{equation}\label{E1}
\tilde\rho_e = div \, {\bf E} = const_x , \qquad {\bf J}_e = curl \, {\bf
B} - {1\over c^2} {\partial {\bf E} \over \partial t} = const_t .
\end{equation}
However, this is a particular case only.
Let me mention that the equation
\begin{equation}\label{1}
{\partial {\bf J}_e \over \partial t} = - grad\, \tilde\rho_e
\end{equation}
follows from (\ref{MY1}) provided that ${\bf J}_e$ and $\tilde\rho_e$
are defined as in Eq. (\ref{E1}).

Now we need to take relations of vector algebra in mind:
\begin{equation}
curl\, curl\, {\bf X} = grad\, div\, {\bf X} -{\bf \nabla}^2 {\bf X},
\end{equation}
where ${\bf X}$ is an arbitrary vector.
Recasting Eq. (\ref{MY1})
and taking the D'Alembert equation (\ref{DA}) in mind
one can come in the general case to
\begin{equation}\label{4}
{\bf J}_m = - {\partial {\bf B} \over \partial t} - curl\, {\bf E} =
grad\,\chi_m,
\end{equation}
in order to satisfy the recasted equation (\ref{MY1})
\begin{equation}
curl \, {\bf J}_m = 0 .
\end{equation}

The second equation (\ref{MY2}) yields
\begin{equation}\label{2}
{\bf J}_e =  curl \, {\bf B} - {1\over c^2}
{\partial {\bf E} \over \partial t} = grad \, \chi_e
\end{equation}
(in order to satisfy $curl\,{\bf J}_e = 0$).
After adding and subtracting
${1\over c^2} \partial^2 {\bf B} / \partial t^2$
one obtains
\begin{equation}\label{EQ2}
\tilde \rho_m = div \, {\bf B} = const_x ,
\qquad{\partial {\bf B} \over \partial t} +curl\, {\bf E} = const_t ,
\end{equation}
provided that
\begin{equation}
{\bf \nabla}^2 {\bf B} -{1\over c^2}
{\partial^2 {\bf B} \over \partial t^2} =0
\end{equation}
(i.~e. again the D'Alembert equation taken into account).
The set of equations (\ref{EQ2}), with the constants are chosen to be zero,
is {\it ``an identity satisfied by certain space-time derivatives
of $F_{\mu\nu}$\ldots, namely,\footnote{Ref.~\cite{FD1,FD2}.}}
\begin{equation}
{\partial F_{\mu\nu} \over \partial x^\sigma} +
{\partial F_{\nu\sigma} \over \partial x^\mu} +
{\partial F_{\sigma\mu} \over \partial x^\nu} = 0.
\end{equation}
However, it is also a particular case.
Again, the general solution is
\begin{eqnarray}\label{3}
{1\over c^2} \frac{\partial {\bf J}_m}{\partial t} = - grad \,
\tilde\rho_m .
\end{eqnarray}

We must pay attention at the universal case.
What are the chi-functions? How should we
name them? From Eqs. (\ref{1}) and (\ref{2}) we conclude
\begin{equation}
\tilde \rho_e = - \frac{\partial
\chi_e}{\partial t} + const,
\end{equation}
and from (\ref{4}) and (\ref{3}),
\begin{equation}
\tilde \rho_m = -
{1\over c^2}\frac{\partial\chi_m}{\partial t} + const,
\end{equation}
what tells us that $\tilde\rho_e$ and $\tilde\rho_m$ are constants provided
that the primary functions $\chi$ are linear functions in time
(decreasing or increasing?).
It is useful to compare the definitions $\tilde\rho_e$ and ${\bf J}_e$
and the fact of an appearance of the functions $\chi$ with the 5-potential
formulation of the electromagnetic theory~\cite{FD2}, see also
ref.~\cite{Gersten},\cite{FD3}-\cite{Malik}.

At last, I would like to note the following. We can obtain
\begin{eqnarray}
div \, {\bf E} &=& 0 , \quad {1\over c^2} {\partial {\bf E} \over \partial t}
- curl \, {\bf B} = 0 ,\\
div\, {\bf B} &=& 0 , \quad
{\partial {\bf B} \over \partial t} + curl\, {\bf E} = 0 ,
\end{eqnarray}
{\it which are just the Maxwell's free-space equations,}
in the definite choice  of the  $\chi_e$ and $\chi_m$, namely,
in the case they are constants. In ref.~\cite{Gersten}
it was mentioned: The solutions of Eqs. (4.21,4.22) of ref.~[1b]
satisfy the equations of the type (\ref{ME0},\ref{ME1}), {\it ``but not
always vice versa".} A interpretation of this statement
and investigations of Eq. (1)  with other initial and boundary conditions
(or of the functions $\chi$) deserve further elaboration (both theoretical
and experimental).

Next, if I use Eq. (\ref{psi})
as field function, of course, the question arises on its
transformation from one to another frame.
I would like to draw your attention
at the remarkable fact which follows from a consideration
of the problem in the momentum representation.
For the first sight, one could conclude that
under a transfer from one to another frame
one has to describe the field by  the Lorentz transformed function
$\psi^\prime ({\bf p}) = \Lambda ({\bf p})\psi ({\bf p})$. However, let us
take into account the possibility of combining the Lorentz, dual (chiral)
and parity transformations in the case of higher spin
equations\footnote{The equations for the four functions $\psi_i^{(k)}$,
Eqs. (8), (10), (18) and (19) of ref.~\cite{DVO1}, reduce to the equations
for ${\bf E}$ and ${\bf B}$, which appear to be the same for each case in
a massless limit.}.  This possibility has been discovered and investigated
in~\cite{DVA3,DVO2}.  The four bispinors $u_1^{\sigma\, (1)} ({\bf p})$,
$u_2^{\sigma\, (1)} ({\bf p})$, $u_1^{\sigma\, (2)} ({\bf p})$ and
$u_2^{\sigma\, (2)} ({\bf p})$, see Eqs. (10), (11), (12) and (13) of
ref.~\cite{DVO3}, form the complete set (as well as $\Lambda ({\bf p})
u_i^{\sigma\, (k)} ({\bf p}))$. Namely,\footnote{After completing the
preliminary version of this article I learnt that equations similar  to
Eq. (\ref{comp}) for the second-type $j=1/2$ and $j=1$ bispinors have been
obtained in ref.~[5b,Eqs.(24,25)]. The equations (22a-23c) of the
above-mentioned reference could also be relevant in the following
discussions.}
\begin{eqnarray} \label{comp}
\lefteqn{a_1 u_1^{\sigma\, (1)} ({\bf p})
\bar u_1^{\sigma\, (1)} ({\bf p}) + a_2 u_2^{\sigma\, (1)} ({\bf p}) \bar
u_2^{\sigma\, (1)} ({\bf p}) +\nonumber}\\
&+& a_3 u_1^{\sigma\, (2)}
({\bf p}) \bar u_1^{\sigma\, (2)} ({\bf p}) + a_4 u_2^{\sigma\, (2)} ({\bf
p}) \bar u_2^{\sigma\, (2)} ({\bf p}) = \openone .
\end{eqnarray}
The constants $a_i$ are defined by the choice of the normalization of
bispinors.  In any other frame we are able to obtain the primary wave
function by choosing the appropriate coefficients $c_i^k$ of the expansion
of the wave function (in fact, by using appropriate dual
rotations and inversions)\footnote{See  Eqs. (17,20) in ref.~\cite{DVO1}.}
\begin{equation}
\Psi = \sum_{i,k=1,2} c_i^k \psi_i^{(k)} .
\end{equation}
Of course, the same statement is valid for
negative-energy solutions, since they coincide with the positive-energy
ones in the case of the Hammer-Tucker formulation for a $j=1$ boson,
ref.~\cite{DVO2,Hammer}. By using the plane-wave expansion
it is easy to prove the validity of the conclusion in the coordinate
representation. Thus, the question of fixing the relative phase
factor by appropriate physical conditions (if
exist) in each point of the space-time
appears to have a physical significance for both massive (charged)
and massless particles in the framework of
relativistic quantum electrodynamics\footnote{The paper
which is devoted to the important experimental consequences of this fact
(e.~g., the Aharonov-Bohm effect and some {\it others}) is in progress.}.

Finally, let me mention that in the nonrelativistic limit $c\rightarrow
\infty$ one obtains the dual
Levi-Leblond's ``Galilean Electrodynamics",
ref.~\cite{Levy,Crawford}.

{\it Conclusion}\footnote{This conclusion also follows from the results
of the paper~\cite{DVA1,DVO1,OLD} and ref.~[1b] provided that
the fact that $({\bf j} {\bf p})$ has no inverse one
has been taken into account.}: The Weinberg-Tucker-Hammer
massless equations (or the Proca equations for
$F_{\mu\nu}$), see also (\ref{ME0}) and (\ref{ME1}), are equivalent to the
Maxwell's equations in the definite choice of the initial and boundary
conditions, what proves their consistency.
They (Eq. (1) for spin $j$) were shown
in ref.~\cite{DVA1} to be free from all kinematical
acausality as opposed to Eqs. (4.21) and (4.22).
Therefore, we have to agree with
Dr. S. Weinberg who spoke out about the equations (4.21)
and (4.22):  {\it ``The fact that these (!) field equations
are of first order for any spin seems
to me to be of no great significance\ldots"}~[1b, p. B888].

Meantime, I would not like to shadow theories based on the use
of the vector potentials, i.~e. of the representation $D(1/2,1/2)$
of the Lorentz group. While the description of the $j=1$ massless
field by using this representation contradicts with the Weinberg theorem
$B-A=\lambda$ one cannot forget about the significant achievements
of these theories. The description proposed here and in my previous
papers~\cite{DVO1}-\cite{DVO4} could be helpful only if we shall
necessitate to go beyond the framework of the Standard Model, i.~e.
if we shall come across the reliable
experiment results which could not have a satisfactory explanation
on the ground of the concept of a minimal coupling~(see, e.~g.,
ref.~\cite{DVA4} for a discussion of
the model which forbids such a form of the interaction).

{\it Acknowledgements.} I thank
Profs. A.~Turbiner and Yu.~F.~Smirnov
for the questions on the non-relativistic limit of
the Weinberg-Tucker-Hammer equations, Prof. I. G. Kaplan,
for useful discussions, and all who gave
a initial impulse to start the work in this direction.

I am grateful to Zacatecas University for a professorship.

In fact, this paper is the {\it Addendum} to the previous
ones~\cite{DVO1}-\cite{DVO3}. It has been thought on September 3-4, 1994
as a result of discussions at the IFUNAM seminar (M\'exico,  2/IX/94).}

\footnotesize{
\vspace*{15mm}
\noindent\hspace*{7cm} Wir sind gew\"ohnt, dass die Menschen\\
\hspace*{7cm} verh\"ohnen was sie nicht verstehen.\\
\medskip
\hspace*{10cm} Goethe\\
\vspace*{15mm}}

\end{document}